\def\ra{\rangle}
\def\la{\langle}
\def\be{\begin{equation}}
\def\ee{\end{equation}}
\def\ba{\begin{array}}
\def\ea{\end{array}}

\def\Cb{{\Bbb C}}

\documentstyle[12pt]{article}
\begin{document}
\input amssym.def
\setcounter{page}{1}
\centerline{\Large\bf Equivalence of Tripartite Quantum  States}
\vspace{2ex} \centerline{\Large\bf under Local Unitary
Transformations} \vspace{4ex}
\begin{center}
Sergio Albeverio$^{a}$ \footnote{SFB 611; IZKS; BiBoS;
CERFIM(Locarno); Acc. Arch. USI (Mendrisio)

~~~e-mail: albeverio@uni-bonn.de}, ~Laura
Cattaneo$^{a}$ \footnote{e-mail: cattaneo@wiener.iam.uni-bonn.de},
 ~Shao-Ming Fei$^{a,b,c}$
\footnote{e-mail: fei@uni-bonn.de}, ~Xiao-Hong
Wang$^{b}$ \footnote{e-mail: wangxh@mail.cnu.edu.cn}

\vspace{2ex}
\begin{minipage}{4.8in}

{\small $~^{a}$ Institut f\"ur Angewandte Mathematik,
Universit\"at Bonn, D-53115}

{\small $~^{b}$ Department of Mathematics, Capital Normal
University, Beijing 100037}

{\small $~^{c}$ Max Planck Institute for
Mathematics in the Sciences, 04103 Leipzig}

\end{minipage}
\end{center}

\vskip 1 true cm
\parindent=18pt
\parskip=6pt
\begin{center}
\begin{minipage}{4.5in}
\vspace{3ex} \centerline{\large Abstract} \vspace{4ex}

The equivalence of tripartite pure states under local
unitary transformations is investigated. The nonlocal properties for a class of
tripartite quantum states in $\Cb^K \otimes \Cb^M \otimes \Cb^N$
composite systems are investigated  and a  complete set of
invariants under local unitary transformations for these states is
presented. It is shown that two of these  states are locally
equivalent if and only if all these invariants have the same
values.

\bigskip
\bigskip

\vfill
\smallskip

\vfill

\end{minipage}
\end{center}

\bigskip
\medskip
\bigskip
\medskip

Quantum entanglement is one of the most striking features of
quantum phenomena \cite{1}.  It is playing very important roles in
quantum information processing such as quantum computation
\cite{DiVincenzo}, quantum teleportation
\cite{teleport} (for discussions of
experimental realizations see \cite{telexp}), dense coding
\cite{dense} and quantum cryptographic schemes
\cite{crypto1}.
As  the degree of entanglement of two parts of a quantum system
remains invariant under local unitary transformations of these
parts,  the invariants of local unitary transformations give rise
to an effective description of entanglement.  Two states are
equivalent under local unitary transformations
if and only if they are assigned the same values by all invariants
under local unitary
transformations. The method developed in \cite{Rains,Grassl}, in
principle, allows one to compute all the invariants of local
unitary transformations, though in general it is not operational.
In \cite{makhlin}, the invariants for general two-qubit systems
are studied and a complete set of 18 polynomial invariants is
presented. It is proven that two qubit mixed states are locally
equivalent if and only if all these 18 invariants have equal
values in these states.  In \cite{linden} three qubits states are
also discussed in detail from a similar point of view.
In \cite{generic} a complete set of invariants is
presented for bipartite generic  mixed states.

In this letter,  we discuss the locally invariant properties of
arbitrary dimensional tripartite quantum states in $\Cb^K \otimes
\Cb^M \otimes \Cb^N$ composite systems. We present a complete set
of invariants for a class  of  pure  states  and show that two of
these  states are locally equivalent if and only if all our
invariants have equal values.

Let $H_A$ resp. $H_B$ resp. $H_C$ be  $K$ resp. $M$ resp. $N$ dimensional
complex Hilbert spaces.  We denote by $\{\vert
e_i\rangle\}_{i=1}^{K}$ , $\{\vert f_i\rangle\}_{i=1}^{M}$  and
$\{\vert h_i\rangle\}_{i=1}^{N}$ the orthonormal bases in $H_A$,
$H_B$ and $H_C$ respectively. A general pure state on $H_A\otimes
H_B\otimes H_C$ is of the form
\begin{equation}\label{mmm}
\vert\Psi\rangle=\sum_{i=1}^K\sum_{j=1}^M\sum_{k=1}^N a_{ijk}\vert
e_i\rangle\otimes \vert f_j\rangle\otimes \vert
h_k\rangle,~~~~~~a_{ijk}\in\Cb
\end{equation}
with the normalization
$\displaystyle\sum_{i=1}^K\sum_{j=1}^M\sum_{k=1}^N
a_{ijk}a_{ijk}^\ast=1$ ($\ast$ denotes complex conjugation).

$|\Psi\ra$ can be regarded as a state on the bipartite systems $A-BC$,
$B-AC$ or $C-AB$. For each such bipartite decomposition, let us
consider the matrix whose entries are the coefficients of the
state $|\Psi\ra$ with respect to the bipartite decomposition. Let
$A_1$ be the matrix corresponding to $|\Psi\ra$ as a bipartite
state in the $A-BC$ system,  with the row  (resp.  column) indices
from  the
 subsystem  A  (resp. BC).  For example, if $K=M=N=2$,
$$ A_1=\left(\ba{cccc} a_{111}&a_{112}&a_{121}&a_{122}\\
a_{211}&a_{212}&a_{221}&a_{222}\\
\ea \right). $$

Similarly, denoting by $A_2$ resp. $A_3$ the matrices treating
 $|\Psi\ra$ as a state in the $B-AC$ resp. $C-AB$ bipartite system,
 for $K=M=N=2$ one has:
  $$ A_2=\left(\ba{cccc} a_{111}&a_{112}&a_{211}&a_{212}\\
a_{121}&a_{122}&a_{221}&a_{222}\\
\ea \right), ~~~~A_3=\left(\ba{cccc} a_{111}&a_{121}&a_{211}&a_{221}\\
a_{112}&a_{122}&a_{212}&a_{222}\\
\ea \right). $$

Taking partial trace of $|\Psi\ra\la\Psi|$ over the respective
subsystems, we have $Tr_1|\Psi\ra\la\Psi|=A^t_1A^*_1$,
 $Tr_2|\Psi\ra\la\Psi|=A^t_2A^*_2$,
 $Tr_3|\Psi\ra\la\Psi|=A^t_3A^*_3$,
 where $t$ represents the transpose of a matrix.
 The  following  quantities  are  invariants
associated with the state $|\Psi\ra$ given by (\ref{mmm}):
 \be\label{invariant} I_{\alpha}
=Tr(Tr_1|\Psi\ra\la\Psi|)^{\alpha},~~~\alpha=1,2,\cdots,S,\ee
where $S=min\{K,M,N\}.$

In fact, if  $|\Psi^\prime\ra=U_1\otimes U_2\otimes U_3|\Psi\ra$,
with $U_i$ unitary matrices acting on the space $H_i$, $i=1,2,3$, then
$A^{\prime}_1$ corresponding to $|\Psi^{\prime}\ra$  and $A_1$
have the following relation:
$$A^{\prime}_1=U_1A_1(U_2\otimes U_3)^t=U_1A_1V^t, $$
where $V=U_2\otimes U_3$ is also a unitary matrix. So we have
$$Tr_1|\Psi^{\prime}\ra\la\Psi^{\prime}|={A^{\prime}}^t_1{A^{\prime}}^*_1
=(u_1A_1V^t)^t(u_1A_1V^t)^*=V(A^t_1A^*_1)V^{\dag} $$
and we get
$$ Tr(Tr_1|\Psi^{\prime}\ra\la\Psi^{\prime}|)^{\alpha}=Tr(V(A^t_1A^*_1)^{\alpha}V^{\dag}) =Tr(A^t_1A^*_1)^{\alpha}
=Tr(Tr_1|\Psi\ra\la\Psi|)^{\alpha},$$
i.e., $I_{\alpha}$, $\alpha=1,\cdots, S$, are invariants.

Similarly, we can construct the following invariants:
\be \label{inv2} J_{\alpha} =Tr(Tr_2|\Psi\ra\la\Psi|)^{\alpha},~~~
\alpha=1,2,\cdots, S ,\ee
\be \label{inv3} K_{\alpha} =Tr(Tr_3|\Psi\ra\la\Psi|)^{\alpha},~~~
\alpha=1,2,\cdots, S.\ee

There are also other invariants like \be
Tr(Tr_i(Tr_j|\Psi\ra\la\Psi|)^{\alpha})^{\beta},~~~ i,j=1,2,3, ~~~
i\not= j,~~\alpha, \beta  =1,2,\cdots, S. \ee Relevant quantities
for states like the Frobenius norm, singular values and the degree
of entanglement are all invariants under local unitary
transformations. Generally one needs all the invariants to judge
whether two tripartite states  are locally  equivalent. However,
for some class of special states, only one kind of invariants, either
(\ref{invariant}), (\ref{inv2}) or (\ref{inv3}), is sufficient, as
we are going to prove. We first recall some results  on  matrix
realignment \cite{chenkai} and give some definitions.

If $Z$ is an $m\times m$ block matrix with each block of size
$n\times n$, the realigned matrix $\tilde{Z}$ is defined by \be
\tilde{Z}=[vec(Z_{11}),\cdots,vec(Z_{m1}),\cdots,
vec(Z_{1m}),\cdots, vec(Z_{mm})]^t,\ee where  $$
vec(A)=[a_{11},\cdots,a_{m1},a_{12},\cdots,a_{m2},\cdots,a_{1n},\cdots,
a_{mn}]^t$$ for  any  $m\times n$ matrix $A$  with  entries
$a_{ij}.$

It is straightforward to verify that a matrix $U$ can
 be expressed as the tensor product of two matrices
 $X$ and $Y,$ i.e. $U=X\otimes Y$,  if  and  only  if
\be \label{ch} \tilde{U}=vec(X)vec(Y)^t\ee
(cf, e.g., \cite{kropro}).

{\sf[Definition].} An $mn\times mn$ unitary matrix $U$ is called
unitarily  decomposable,  if  there  exist an $m\times m$  unitary
matrix $U_1$ and an $n\times n$ unitary matrix $U_2$, such that
$U=U_1\otimes U_2$.

{\sf[Lemma ].} Let $U$ be an $mn\times mn$ unitary matrix. $U$ is a unitarily
decomposable matrix if and only if the rank of $\tilde{U}$ is one,  $r(\tilde{U})=1$.

{\sf[Proof].} Let $U$ be a unitarily decomposable matrix, i.e.,
there exist unitary matrices $U_1$ and $U_2$ such that
$U=U_1\otimes U_2$. Applying (\ref{ch}) and using the property
that a matrix is rank one if and only if it can be written as
product of a column vector and a row vector, we have
$r(\tilde{U})=1.$

Conversely, if $r(\tilde{U})=1$, there  are  matrices $X$ and $Y$
such  that  $U=X\otimes Y$. On the other hand, due to the
unitarity of $U,$ $X$ and $Y$ should satisfy the following
equation:
$$ UU^{\dag}=(X\otimes Y)(X^{\dag}\otimes
Y^{\dag}) =XX^{\dag}\otimes YY^{\dag}=I_{mn}.$$
Let $x_{ij}$
denote the entries of $XX^{\dag}$. The above relation
implies that $x_{ij}=0$ if $i\not=j$ and $x_{ii}=k^{-1}\not=0,~~i,j=1,\cdots,
m,$ and $YY^{\dag}$ is a diagonal scalar matrix, i.e.
$XX^{\dag}=k^{-1}I_m$ and $YY^{\dag}=kI_n$.

Similarly, we have $X^{\dag}X=k'^{-1}I_m,$ and $Y^{\dag}Y=k'I_n.$
It is easily proven that $k'=k$. Therefore
$XX^{\dag}=X^{\dag}X=k^{-1}I_m$ and $YY^{\dag}=Y^{\dag}Y=kI_n.$
Since $XX^{\dag}$ and $YY^{\dag}$ are positive and selfadjoint,
$k$ is real and positive. Hence $U_1=\sqrt{k}X$ and
$U_2=\frac{Y}{\sqrt{k}}$ are unitary matrices such that $U=U_1\otimes U_2$ is unitarily
decomposable.  ~~~~~~~~~~~~~~~~ $\Box$

Note that if
$U=X\otimes Y$ is a unitary  matrix, then $X$ and $Y$ are either
both  unitary or both not unitary.

We can judge whether an $mn\times mn$ unitary matrix $U$ is
unitarily decomposable or not in the following way:
if the rank of the realigned matrix $\tilde{U}$ is not one,
$r(\tilde{U})\not=1$, then $U$ is not decomposable. If
$r(\tilde{U})=1$, then it can be written as a product of a column
vector and a row vector, i.e., there  exist $(a_1,\cdots,
a_{m^2})^t$  and $(b_1,\cdots, b_{n^2})$ such that
 $\tilde{U}=(a_1,\cdots,a_{m^2})^t(b_1,\cdots,b_{n^2}).$
These vectors can be obtained from the realignment  of certain
 matrices, say $vec(X)=(a_1,\cdots, a_{m^2})^t,$
$vec(Y)=(b_1,\cdots, b_{n^2})^t,$ so that $U=X\otimes Y.$ If one
of $X$ and $Y$ is unitary, then $U$ is unitarily decomposable.

We consider now the state $|\Psi\ra$ in (\ref{mmm}) as a bipartite
state $A-BC$. As shown in \cite{generic}, two bipartite states
$|\psi\ra=\sum_{{i}=1}^M\sum_{{j}=1}^N a_{ij}|ij\ra$ and
$|\psi^{\prime}\ra=\sum_{{i}=1}^M\sum_{{j}=1}^N a^{\prime}_{ij}|ij\ra$ are
equivalent under local unitary transformations
if and only if they are assigned the same values for all the invariants:
$T_{\alpha}=T^{\prime}_{\alpha}$, for $\alpha=1,\cdots, min\{N,M\},$
where $T_{\alpha}=Tr(AA^{\dag})^{\alpha},$
$T^{\prime}_{\alpha}=Tr({A^{\prime}}{A^{\prime}}^{\dag})^{\alpha},$
and $A$, $A^{\prime}$ are the $M\times N$ matrices with the
entries $a_{ij}$ and $a^{\prime}_{ij}$ respectively.  If $T_{\alpha}
=T^{\prime}_{\alpha},$ there exist unitary matrices $U$ and $V$
such that $|\psi^{\prime}\ra=U\otimes V |\psi\ra $, which also
implies $A^{\prime}=UAV^t$, i.e., $ AA^{\dag}$ and
$A^{\prime}{A^{\prime}}^{\dag}$ are unitary equivalent and have the same singular values. $U$ and
$V$ are dependent on $|\psi\ra$  and $|\psi^{\prime}\ra,$ and
can be obtained by using the singular value decomposition method: $U=u'u^{\dag}$
and $V=v'v^{\dag}$, where $A=uDv^{\dag}$ and $A'=u'Dv'^{\dag}$ are singular value
decompositions of $A$ and $A'$, respectively, with the singular values ordered descending.

Summarizing the above discussions we have the following theorem:

{\sf[Theorem].} If two tripartite states $|\Psi\rangle$ and $|\Psi'\rangle$ on
    $H_A\otimes H_B\otimes H_C$ have the same values of the
    invariants given by (2), i.e. $I_\alpha=I_\alpha'$ for
    $\alpha=1,\dots,S$, there are unitary matrices $U_1$
on $H_A$ and $V_1$ on $H_B\otimes H_C$ such that
$|\Psi^{\prime}\ra=U_1\otimes V_1|\Psi\ra$. $|\Psi\ra$ and
$|\Psi^{\prime}\ra$ are then equivalent under local unitary
transformations if $V_1$ satisfies $r(\tilde{V_1})=1$.

{\sf[Remark].} If we
say that two pure tripartite states $|\Psi\ra$
and $|\Psi^{\prime}\ra$ are a pair of $D_1$ states if
they satisfy $|\Psi^{\prime}\ra=U_1\otimes V_1|\Psi\ra$ with
$U_1$ a unitary matrix on $H_A$ and $V_1$ a unitarily decomposable matrix
on $H_B\otimes H_C$, we have defined an equivalence relation
$|\Psi^{\prime}\ra \sim |\Psi\ra$. Indeed, as
$|\Psi\ra=U^{\dag}_1\otimes V^{\dag}_1 |\Psi^{\prime}\ra$, where
$U^{\dag}_1$ is unitary, and $V^{\dag}_1$ is also unitarily decomposable
with $r(\tilde{V}^{\dag}_1)=1$, one has that if $|\Psi^{\prime}\ra \sim |\Psi\ra$
then $|\Psi\ra \sim |\Psi^{\prime}\ra$.
Transitivity also holds, namely, if $|\Psi^{\prime\prime}\ra \sim |\Psi^{\prime}\ra$ and $|\Psi^{\prime}\ra \sim |\Psi\ra$, then
$|\Psi^{\prime\prime}\ra \sim |\Psi\ra$.

We shall provide two examples to illustrate our results.

{\sf[Example 1].} We consider two states
$|\Psi\ra=\frac{1}{\sqrt{2}}(|001\ra+|100\ra),$ and
$|\Psi^{\prime}\ra=\frac{1}{\sqrt{2}}(|010\ra+|111\ra)$ in
$H_A\otimes H_B\otimes H_C, $ where $K={\rm dim}H_A=2, M={\rm
dim}H_B=2, N={\rm dim}H_C=2$. Let us denote by $\{|0\ra,~|1\ra\}$
the orthonormal basis of $H_A$, $H_B$, and $H_C$. We have
$$\rho=Tr_1|\Psi\ra\la\Psi|=\rm
diag(\frac{1}{2},\frac{1}{2},0,0),~~~
\rho^{\prime}=Tr_1|\Psi^{\prime}\ra\la\Psi^{\prime}|= \rm
diag(0,0,\frac{1}{2},\frac{1}{2}),$$ and
$$I_{\alpha}=Tr(Tr_1|\Psi\ra\la\Psi|)^{\alpha}=\frac{1}{2^{\alpha-1}},~~~~
I^{\prime}_{\alpha}=Tr(Tr_1|\Psi^{\prime}\ra\la\Psi^{\prime}|)^{\alpha}
=\frac{1}{2^{\alpha-1}}.$$

Since $I_{\alpha}=I^{\prime}_{\alpha}$, we treat $|\Psi\ra$ and
$|\Psi^{\prime}\ra$ as states in the bipartite system $H_A\otimes
H_{BC}$, where $H_{BC}=H_B\otimes H_C.$ Then we get the
corresponding $2\times 2$ block matrices $A_1=\left(\ba{cc}
T_1&0\\0&0\ea\right),$ $A^{\prime}_1=\left(\ba{cc}
0&T^{\prime}_1\\0&0\ea\right),$ where $T_1=\left(\ba{cc}
0&\frac{1}{\sqrt{2}}\\\frac{1}{\sqrt{2}}&0\ea\right),$
$T^{\prime}_1=\left(\ba{cc} \frac{1}{\sqrt{2}}&0\\0
&\frac{1}{\sqrt{2}}\ea\right).$ From the singular value
decomposition of matrices we have unitary matrices $U_1$ in $H_A$
and $V_1$ in $H_B\otimes H_C$ such that
$|\Psi^{\prime}\ra=U_1\otimes V_1|\Psi\ra$. In this case $V_1=I$.
Therefore $|\Psi\ra$ and $|\Psi^{\prime}\ra$ are $D_1$ states and
they are equivalent under local unitary transformations.

{\sf[Example 2].} We consider two states
$|\Psi\ra=\frac{1}{\sqrt{2}}(|110\ra+|012\ra),$ and
$|\Psi^{\prime}\ra=-\frac{\sqrt{6}}{4}|000\ra
+\frac{\sqrt{2}}{4}|010\ra
-\frac{\sqrt{3}}{4}|101\ra+\frac{\sqrt{3}}{4}|102\ra
+\frac{1}{4}|111\ra-\frac{1}{4}|112\ra$ in $H_A\otimes H_B\otimes
H_C, $ where $K={\rm dim}H_A=2, M={\rm dim}H_B=2, N={\rm
dim}H_C=3$. Let us denote by $\{|0\ra,~|1\ra\}$ the orthonormal
basis of $H_A$ and $H_B$ , and by $\{|0\ra,~|1\ra,~|2\ra\}$ the
orthonormal basis of $H_C$. We have
$$I_{\alpha}=Tr(Tr_1|\Psi\ra\la\Psi|)^{\alpha}=\frac{1}{2^{\alpha-1}},~~~~
I^{\prime}_{\alpha}=Tr(Tr_1|\Psi^{\prime}\ra\la\Psi^{\prime}|)^{\alpha}
=\frac{1}{2^{\alpha-1}}.$$

Since $I_{\alpha}=I^{\prime}_{\alpha}$, we treat $|\Psi\ra$ and
$|\Psi^{\prime}\ra$ as states in the bipartite system $H_A\otimes
H_{BC}$, where $H_{BC}=H_B\otimes H_C;$ the corresponding $2\times
6$  matrices $A_1$ and $A_1'$ are, respectively,
$\left(\ba{cccccc}
0&0&0&0&0&\frac{{\sqrt{2}}}{2}\\0&0&0&\frac{{\sqrt{2}}}{2}&0&0\ea\right)$
and
$\left(\ba{cccccc}-\frac{\sqrt{6}}{4}&0&0&\frac{\sqrt{2}}{4}&0&0\\
0&-\frac{\sqrt{3}}{4}&\frac{\sqrt{3}}{4}&0&\frac{1}{4}&-\frac{1}{4}
\ea\right).$ The singular value decomposition delivers us
unitary matrices $U_1$ in $H_A$ and $V_1$ in $H_B\otimes H_C$ such
that $|\Psi^{\prime}\ra=U_1\otimes V_1|\Psi\ra$. For instance,
$$U_1=\left(\ba{cc} 0&-1\\1&0\\ \ea\right)\quad\textrm{and}\quad V_1=\left(\ba{cccccc}
{1}/{2}&0&0&{\sqrt{3}}/{2}&0&0\\
0&{\sqrt{2}}/{4}&-{\sqrt{2}}/{4}&0&{\sqrt{6}}/{4}&{{-\sqrt{6}}}/{4}\\
0&{{\sqrt{2}}}/{4}&{{\sqrt{2}}}/{4}&0&{{\sqrt{6}}}/{4}&{{\sqrt{6}}}/{4}\\
{{\sqrt{3}}}/{2}&0&0&-{1}/{2}&0&0\\
0&{{\sqrt{6}}}/{4}&-{{\sqrt{6}}}/{4}&0&-{{\sqrt{2}}}/{4}&{{\sqrt{2}}}/{4}\\
0&{{\sqrt{6}}}/{4}&{{\sqrt{6}}}/{4}&0&-{{\sqrt{2}}}/{4}&-{{\sqrt{2}}}/{4}\\
\ea\right).$$ The rank of $\tilde{V_1}$ is one, therefore
$|\Psi\ra$ and $|\Psi^{\prime}\ra$ are $D_1$ states and they are
equivalent under local unitary transformations.

{\sf[Remark].} We can also say that
two pure tripartite states $|\Psi\ra$
and $|\Psi^{\prime}\ra$ are a pair of $D_2$ (resp. $D_3$) states.
For example, if we treat $|\Psi\ra$ as a state in the B-AC system,
then $Tr_2|\Psi\ra\la\Psi|=A^t_2A^*_2$. If
$J_{\alpha}=J^{\prime}_{\alpha}$, from the result on bipartite
systems we have that $|\Psi^{\prime}\ra=U_2\otimes V_2|\Psi\ra$,
where $U_2$ acts on $H_B$ and $V_2$ on $H_A\otimes H_C$. If the
unitary  matrix $V_2$ satisfies $r(\tilde{V_2})=1$, then $|\Psi\ra$
and $|\Psi^{\prime}\ra$ are a pair of $D_2$ states
and they are equivalent under local unitary transformations.
A pair of $D_3$ states can be defined in a similar way.

If $|\Psi\ra$ and  $|\Psi^{\prime}\ra$ are not a pair of $D_1$ states, one
can check whether they are a pair of  $D_2$ or $D_3$ states, by using
$J_{\alpha}$ and $J^{\prime}_{\alpha}$ or $K_{\alpha}$ and
$K^{\prime}_{\alpha}$ to check whether $|\Psi\ra$ and
$|\Psi^{\prime}\ra$ are equivalent under local
unitary transformations or not.

In summary, we have discussed the local invariants for arbitrary
dimensional tripartite quantum states in $\Cb^K \otimes \Cb^M
\otimes \Cb^N$ composite systems and have presented a set of
invariants under local unitary transformations. The
invariants are not necessarily independent (they could be
represented by each other in some cases), but the invariants are sufficient to
judge whether two states constitute a pair of $D_i$, $i=1,2,3$, states, which
are equivalent under local unitary transformations.

\vspace{1.0truecm}

\noindent {\bf Acknowledgments} The second named author gratefully
acknowledges the financial support by the Stefano Franscini Fund.
The fourth author gratefully acknowledges the support provided by
the China-Germany Cooperation Project 446 CHV 113/231, ``Quantum
information and related mathematical problems".


\vspace{1.0truecm}


\begin{thebibliography}{99}
\bibitem{1} A. Peres, Quantum Mechanics: Concepts and Methods, Kluwer, Dordrecht (1993).

\bibitem{DiVincenzo} See, for example, D.P. Di Vincenzo,
{\it Science} {\bf 270}, 255 (1995);\\
M. Nielsen and I.L. Chuang, {\it Quantum Computation and
Quantum Information}, Cambridge University Press (2000).

\bibitem{teleport} C.H. Bennett, G. Brassard, C. Cr\'epeau,
       R. Jozsa, A. Peres,
       and W.K. Wootters, {\it Phys. Rev. Lett.} {\bf 70}, 1895 (1993);\\
S. Albeverio and S.M. Fei,
          {\it Phys. Lett. A} {\bf 276}, 8-11 (2000);\\
G.M. D'Ariano, P. Lo Presti, and M.F. Sacchi, {\it Phys. Lett. A} {\bf
272}, 32 (2000);\\
S. Albeverio, S.M. Fei, and W.L. Yang,
{\em Phys. Rev. A} {\bf 66}, 012301 (2002).

\bibitem{telexp}
D. Bouwmeester, J.-W. Pan, K. Mattle, M.
Eibl, H. Weinfurter, and A. Zeilinger, {\it Nature} {\bf 390}, 575 (1997);\\
D. Boschi, S. Brance,
F. De Martini, L. Hardy, and S. Popescu, {\it Phys. Rev. Lett.} {\bf 80}, 1121 (1998);\\
A. Furusawa, J.L. S{\o}rensen, S.L. Braunstein, C.A. Fuchs, H.J.
Kimble, and
E.S. Polzik, {\it Science} {\bf 282}, 706 (1998);\\
M.A. Nielsen, E. Knill, and R. Laflamme, {\it Nature} {\bf 396}, 52 (1998).

\bibitem{dense} C.H. Bennett and S.J. Wiesner,
        {\it Phys. Rev. Lett.} {\bf 69}, 2881 (1992).

\bibitem{crypto1}
A. Ekert, {\it Phys. Rev. Lett.} {\bf 67}, 661 (1991);\\
D. Deutsch, A. Ekert, P. Rozsa, C. Macchiavello,
S. Popescu, and A. Sanpera, {\it Phys. Rev. Lett.} {\bf 77}, 2818 (1996);\\
C.A. Fuchs, N. Gisin, R.B. Griffiths, C-S. Niu, and A. Peres, {\it
Phys. Rev. A} {\bf 56}, 1163 (1997).

\bibitem{Rains}
E.M. Rains, {\it IEEE Transactions on Information Theory} {\bf 46}, 54 (2000).

\bibitem{Grassl}
M. Grassl, M. R\"otteler, and T. Beth, {\it Phys. Rev. A} {\bf 58}, 1833
(1998).

\bibitem{makhlin}
Y. Makhlin, {\it Quant. Info. Proc.} {\bf 1}, 243 (2002).

\bibitem{linden}
N. Linden, S. Popescu, and A. Sudbery, {\it Phys. Rev. Lett.} {\bf
83}, 243 (1999).

\bibitem{generic}
S. Albeverio, S.M. Fei, P. Parashar, and W.L. Yang,
{\it Phys. Rev. A} {\bf 68}, 010313 (R) (2003).

\bibitem{chenkai} K. Chen and L.A. Wu, {\it Phys. Lett. A} {\bf 306}, 14 (2002).

\bibitem{kropro} N.P. Pitsianis, Ph.D. thesis, {\it The Kronecker
Product in Approximation and Fast Transform Generation}, Cornell
University, New York (1997).

\end{thebibliography}
\end{document}